\documentclass[a4paper,12pt]{article}
\usepackage[T2A]{fontenc} 
\usepackage[russian, english]{babel} 
\usepackage[dvips]{graphicx} 
\makeindex 
\usepackage{epsfig} 
\usepackage[dvips]{texdraw} \input{txdtools}
\usepackage{picinpar} 
\usepackage{amsmath}
\usepackage{amssymb}
\usepackage{indentfirst} 

\begin{document}
\begin{center}

{\bf \large Upper Limits on Electric and Weak Dipole Moments of $W$-Boson}

\vspace{0.5cm} 

A.E.\,Blinov\footnote{A.E.Blinov@inp.nsk.su} and
A.S.\,Rudenko\footnote{A.S.Rudenko@inp.nsk.su}

\vspace{0.5cm}

Budker Institute of Nuclear Physics,\\ 630090 Novosibirsk, Russia,\\ and Novosibirsk
State University

\end{center}

\begin{abstract}
The total cross-sections of the reaction $e^+e^-\to W^+W^-$, as measured at LEP-II
at centre-of-mass energies between 183 and 207 GeV are used to derive the upper
limits on the parameters of $CP$-violating ($P$-odd and $C$-even) triple gauge-boson
couplings $WW\gamma$ and $WWZ$. The 95\% CL limits $|\widetilde{\kappa}_Z|<0.13$ and
$|\widetilde{\lambda}_Z|<0.31$ are obtained assuming local $SU(2)_L\times U(1)_Y$
gauge invariance which dictates the relations
$\widetilde{\kappa}_Z=-\tan^2\theta_w\,\widetilde{\kappa}_\gamma$,
$\widetilde{\lambda}_Z=\widetilde{\lambda}_\gamma$. Our results are comparable with
the previous ones obtained through the analysis of the $W$ decay products.

We also discuss the upper limits on the electric dipole moment (EDM) of the
$W$-boson, which follow from the precision measurements of the electron and neutron
EDM.
\end{abstract}

\subsection*{\normalsize 1. Introduction}

The existence of the electric dipole moments and weak dipole moments of elementary
particles would imply $CP$ violation. Since the expected values of $CP$-odd dipole
moments in the Standard Model (SM) are extremely small the measurement of
significantly larger values would be evidence for physics beyond the SM.

The limits on $CP$-odd dipole moments of $W$-boson can be derived both from the
analysis of high-energy experiments and from strict experimental bounds on the EDMs
of other particles.

$CP$-violating triple gauge-boson couplings (TGC) $WW\gamma$ and $WWZ$ in the
high-energy $e^+e^-$ collisions are investigated in
Refs.~\cite{gp,lvy,ckk,dnn,opal,del} by constructing observables sensitive to these
couplings. However, one may obtain quite strict limits on the TGC parameters from
the analysis of the total cross-sections only.

The most strict bounds on electric dipole moments of elementary particles are given
in Table~\ref{table:1}.
\begin{table*}[h]
\begin{center}
\begin{tabular}{|c|c|c|c|c|} \hline
& & & &  \\
& $e$ & $n$ & $p$ & $\mu$  \\
& & & &  \\ \hline
& & & &  \\
$d/e$, cm & $<1.6\times 10^{-27}$ \cite{bcr} & $<2.9\times 10^{-26}$ \cite{bak} &
$<0.79\times
10^{-24}$ \cite{ds,for} & $<1.8\times 10^{-19}$ \cite{ben}\\
& & & & \\ \hline
\end{tabular}
\caption{The upper limits on electric dipole moments of elementary particles}
\label{table:1}
\end{center}
\end{table*}

\vspace{-2mm}

Contributions to the EDMs presented in Table~\ref{table:1} can originate both from
the electric dipole and magnetic quadrupole moments of the $W$-boson.

The outline of the paper is as follows. In section 2 we consider $e^+e^-$
annihilation into $W^+W^-$ and obtain the bounds on the $CP$-violating TGC
parameters from the analysis of the total cross-sections. In section 3 we derive
upper limits on the EDM of the $W$-boson from those on the electron and neutron EDM.

\subsection*{\normalsize 2. The dipole moments of the $W$-boson and $e^+e^-$ annihilation}

\begin{figure}[h]
\center
\begin{tabular}{c c}
\includegraphics[scale=1.2]{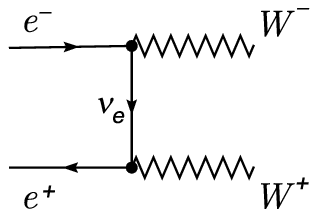} &
\includegraphics[scale=1.2]{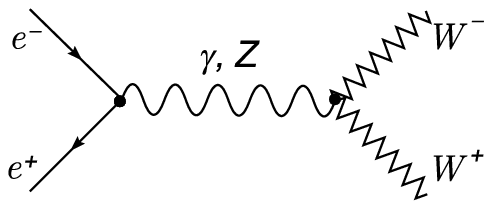}
\end{tabular}
\caption {The Standard Model diagrams} \label{fig:1}
\end{figure}

We start with the consideration of the high-energy electron-positron annihilation
into $W^+W^-$. In the Standard Model this reaction at the tree level is described by
the Feynman graphs shown in Fig.~\ref{fig:1}.

\begin{figure}[h]
\center
\includegraphics[scale=1.2]{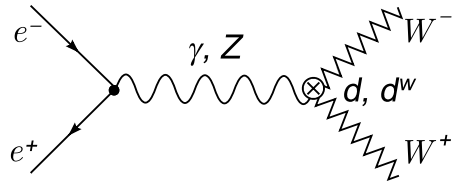}
\caption {The diagrams with the dipole moments vertices} \label{fig:2}
\end{figure}

The diagrams with the intermediate $\gamma$ and $Z$ contain three-vector-boson
vertices $WW\gamma$ and $WWZ$. The general parametrization describing the vertex
$WWV$ (here $V$ means $\gamma$ or $Z$) includes 14 couplings, 7 for $V=\gamma$ and 7
for $V=Z$: $g^V_1$, $\kappa_V$, $\lambda_V$, $g^V_4$, $g^V_5$,
$\widetilde{\kappa}_V$ and $\widetilde{\lambda}_V$ \cite{hag,gou}. We restrict
ourselves to considering only $P$-odd and $C$-even couplings, namely
$\widetilde{\kappa}_V$ and $\widetilde{\lambda}_V$, because only these ones are
related to the electric dipole moment $d$ and weak dipole moment (WDM) $d^w$ of the
$W$-boson (see the corresponding Feynman graphs at the tree level in
Fig.~\ref{fig:2}).

We use the following effective Lagrangian describing interaction of the $W$-boson
EDM with the electromagnetic field and the WDM with the $Z$-boson field:
\begin{equation} \label{lagr}
L_{WWV}=ie_V\widetilde{\kappa}_V W^\dagger_\mu W_\nu
\widetilde{V}^{\mu\nu}+\frac{ie_V\widetilde{\lambda}_V}{m^2_W}\left(\widetilde{W}^\dagger_{\mu\nu}
W^\nu{}_\rho V^{\rho\mu} + W^\dagger_{\mu\nu} \widetilde{W}^\nu{}_\rho V^{\rho\mu} -
W^\dagger_{\mu\nu} W^\nu{}_\rho \widetilde{V}^{\rho\mu}\right),
\end{equation}

\noindent here $e_\gamma=e$, $e_Z=e\cot\theta_w$, where $e$ is the electron charge,
$\theta_w$ is the Weinberg angle;

\vspace{1mm} \noindent $W_\mu$ is the $W^-$ field,

\vspace{1mm} \noindent $V_\mu$ is the photon or the $Z$-boson field, corresponding
to $V=\gamma$ or $V=Z$, respectively;

\vspace{1mm} \noindent $W_{\mu\nu}=\partial_\mu W_\nu - \partial_\nu W_\mu$,\quad
$V_{\mu\nu}=\partial_\mu V_\nu- \partial_\nu V_\mu$;

\vspace{1mm} \noindent
$\widetilde{W}^{\mu\nu}=\frac{1}{2}\epsilon^{\mu\nu\rho\sigma}
W_{\rho\sigma}$,\quad $\widetilde{V}^{\mu\nu}=\frac{1}{2}\epsilon^{\mu\nu\rho\sigma} V_{\rho\sigma}$.\\

It should be noted that the choice of the dimension-6 operator in (\ref{lagr}) is
not unambiguous and sometimes this operator is chosen in a simpler form:
\begin{equation*} \label{lam}
\frac{ie_V\widetilde{\lambda}_V^\prime}{m^2_W}W^\dagger_{\mu\nu}
W^\nu{}_\rho\widetilde{V}^{\rho\mu}.
\end{equation*}
However, this discrepancy leads only to a different $q^2$ dependence of the
form-factors. Since the energy region covered by LEP-II is not wide, one may take
the form-factors to be approximately constant. Therefore, the form of the
dimension-6 operator is not significant to the problem under consideration.

\vspace{2mm}

\begin{figure}[h]
\center
\includegraphics[scale=1.2]{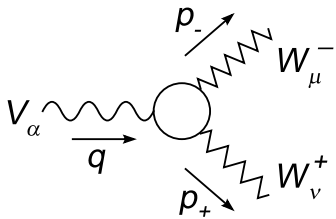}
\caption {The $WWV$ vertex} \label{fig:3}
\end{figure}

The $W^-_\mu(p_{-})W^+_\nu(p_{+})V_\alpha(q)$ vertex (see Fig.~\ref{fig:3})
corresponding to the Lagrangian (\ref{lagr}) for on-shell $W$-bosons looks as
follows \cite{hag}:
\begin{equation} \label{ver}
ie_V\Gamma^{\mu\nu\alpha}_V (p_{-}, p_{+}, q)=ie_V\left(-f^V_6
\epsilon^{\alpha\mu\nu\rho}q_\rho - \frac{f^V_7}{m^2_W}(p_{-}-p_{+})^\alpha
\epsilon^{\mu\nu\rho\sigma}q_\rho (p_{-}-p_{+})_\sigma\right),
\end{equation}

\noindent here $f^V_6=\widetilde{\kappa}_V-\widetilde{\lambda}_V$,
$f^V_7=-\frac{1}{2}\widetilde{\lambda}_V$.

The couplings $\widetilde{\kappa}_V$ and $\widetilde{\lambda}_V$ are related to the
$W$-boson EDM and WDM by \cite{hag,quei}:
\begin{equation} \label{d}
d=\frac{e}{2m_W}\left(\widetilde{\kappa}_\gamma+\widetilde{\lambda}_\gamma\right),
\end{equation}

\begin{equation} \label{dw}
d^w=\frac{e_Z}{2m_W}\left(\widetilde{\kappa}_Z+\widetilde{\lambda}_Z\right).
\end{equation}

$W$-bosons in the reaction $e^+e^- \to W^+W^-$ are produced in the states with the
spin $S=0, 2$ if the production mechanism is regular (because $CP=+1$ and
$CP=(-1)^S$). However, if produced via the $CP$-odd vertex, $W$-bosons will be in
the triplet states $S=1$ ($CP=-1$ and $CP=(-1)^S$). Therefore, if polarization of
the particles is not taken into account the dipole moment vertices do not interfere
with the regular ones, and hence their contribution to the cross-section is of
second order in the dipole moments. Nevertheless the most strict limits on the
dipole moments may be obtained from the analysis of the total cross-sections due to
larger statistics. In this way the best restrictions on the dipole moments of
$b$-quark are derived from the total cross-sections measured at LEP for $e^+e^-$
annihilation into $q\bar{q}$ \cite{br}.

We consider the reaction $e^+e^-\to W^+W^-$ in the centre-of-mass system assuming
unpolarized electron and positron beams, neglecting the electron mass and summing
over polarization of the final particles. Using the $WWV$ vertex parametrization
(\ref{ver}) we obtain the following expressions for the squared matrix elements
$|M|^2_{f^V_i}$ and the interference terms $2\mathrm{Re}\left(M^\dagger_{f^V_6}
M_{f^V_7}\right)$ and $2\mathrm{Re}\left(M^\dagger_{f^\gamma_i} M_{f^Z_j}\right)$:

\begin{equation}
|M|^2_{f^V_i}=C^V |f^V_i|^2 F_{ii},\ \mathrm{where}\ V=\gamma, Z\ \mathrm{and}\
i=6,7;
\end{equation}

\begin{equation}
2\mathrm{Re}\left(M^\dagger_{f^V_6} M_{f^V_7}\right)=C^V 2\mathrm{Re}\left(f^{V *}_6
f^V_7 \right)F_{67},\ \mathrm{where}\ V=\gamma, Z;
\end{equation}

\begin{equation}
2\mathrm{Re}\left(M^\dagger_{f^\gamma_i} M_{f^Z_j}\right)=C^{\gamma Z}
2\mathrm{Re}\left(f^{\gamma *}_i f^Z_j \right) F_{ij},\ \mathrm{where}\ i,j=6,7.
\end{equation}

The functions $F_{ij}$ are as follows:

\begin{equation}
F_{66}=e^4\frac{E^2}{m^2_W}\left(1+\cos^2\theta+\frac{m^2_W}{E^2}\sin^2\theta\right),
\end{equation}

\begin{equation}
F_{77}=16e^4\frac{E^4}{m^4_W}\left(1-\frac{m^2_W}{E^2}\right)^2\sin^2\theta,
\end{equation}

\begin{equation}
F_{67}=F_{76}=4e^4\frac{E^2}{m^2_W}\left(1-\frac{m^2_W}{E^2}\right)\sin^2\theta,
\end{equation}

\noindent where $E$ is the beam energy.

\vspace{2mm}

The functions $C^V$ and $C^{\gamma Z}$ are:
\begin{equation}
C^\gamma=1,
\end{equation}

\begin{equation}
C^Z=\left[\frac{4E^4}{(4E^2-m^2_Z)^2}\frac{V^2_e+A^2_e}{\sin^4\theta_w}\right],
\end{equation}

\begin{equation}
C^{\gamma Z}=\left[-\frac{2E^2}{4E^2-m^2_Z}\frac{V_e}{\sin^2\theta_w}\right],
\end{equation}

\noindent where $V_e=-\frac{1}{2}+2\sin^2 \theta_w$, $A_e=-\frac{1}{2}$.

\vspace{2mm}

As common, we suppose hereafter the form-factors $f^V_6$ and $f^V_7$ being real
\cite{hag}. We assume also local $SU(2)_L\times U(1)_Y$ gauge invariance to be
preserved. It dictates \cite{bil,hwz}:
\begin{equation} \label{kappa}
\widetilde{\kappa}_Z=-\tan^2\theta_w\,\widetilde{\kappa}_\gamma\,,
\end{equation}
\begin{equation} \label{lambda}
\widetilde{\lambda}_Z=\widetilde{\lambda}_\gamma\,.
\end{equation}

So the total cross-section of the process $e^+e^-\to W^+W^-$ looks as follows:
\begin{multline} \label{cs}
\sigma=\sigma_{SM}+ \frac{1}{32\pi}\frac{1}{4E^2}\sqrt{1-\frac{m_W^2}{E^2}}\times \\
\times\int \left(\sum_{\substack{V=\gamma, Z\\i=6,7}}C^V \left(f^V_i\right)^2
F_{ii}+2\sum_{V=\gamma, Z}C^V\left(f^{V}_6 f^V_7
\right)F_{67}+2\sum_{\substack{i=6,7\\j=6,7}}C^{\gamma Z}\left(f^{\gamma}_i f^Z_j
\right) F_{ij}\right) d(\cos\theta).
\end{multline}

To derive the limits, we use the LEP-II measurements of the $W$-pair production
cross-section at centre-of-mass energies between 183 and 207 GeV combined by
\linebreak LEPEWWG \cite{LEPEWWG}. The measured values and the SM predictions of
$e^+e^-\to W^+W^-$ cross-sections are obtained from Table~5.1 and Table~5.2 of
Ref.~\cite{LEPEWWG}, respectively. The residuals were fitted by the second term of
formula (\ref{cs}) with non-zero $\widetilde{\kappa}_Z$ and $\widetilde{\lambda}_Z$
assuming the relations (\ref{kappa}, \ref{lambda}). The correlation matrix for the
LEP combined W-pair cross-sections (Table~A.2 of Ref.~\cite{LEPEWWG}) has been taken
into account. The fit yields the 68\% CL limits $|\widetilde{\kappa}_Z|<0.07$,
$|\widetilde{\lambda}_Z|<0.18$ and the 95\% CL limits $|\widetilde{\kappa}_Z|<0.13$,
$|\widetilde{\lambda}_Z|<0.31$.

The values of $\widetilde{\kappa}_Z$ and $\widetilde{\lambda}_Z$ obtained from
$e^+e^-$ annihilation are presented in Table~\ref{table:2}. Our results obtained
from the analysis of the total cross-sections are comparable with those of
Refs.~\cite{opal,del} derived through analysis of the $W$ production angular
distribution and distributions of the $W$ decay products.

\begin{table*}[h]
\begin{center}
\renewcommand{\arraystretch}{1.3}
\begin{tabular}{|c|c|c|} \hline
& $\widetilde{\kappa}_Z$ & $\widetilde{\lambda}_Z$ \\
\hline
OPAL (2001) \cite{opal} & $-0.20^{+0.10}_{-0.07}$ & $-0.18^{+0.24}_{-0.16}$ \\
($2E=189$ GeV) & & \\
\hline
DELPHI (2008) \cite{del} & $-0.09^{+0.08}_{-0.05}$ & $-0.08 \pm 0.07$ \\
($2E=189-209$ GeV) & & \\
\hline
This work & $|\widetilde{\kappa}_Z|<0.07$ ($68\%$ CL) & $|\widetilde{\lambda}_Z|<0.18$ $(68\%$ CL) \\
($2E=183-207$ GeV) & $|\widetilde{\kappa}_Z|<0.13$ ($95\%$ CL) & $|\widetilde{\lambda}_Z|<0.31$ $(95\%$ CL) \\
\hline
\end{tabular}
\caption{The limits on $\widetilde{\kappa}_Z$ and $\widetilde{\lambda}_Z$ obtained
from $e^+e^-$ annihilation} \label{table:2}
\end{center}
\end{table*}

\subsection*{\normalsize 3. The $W$-boson EDM from the electron and neutron EDM}

In this section we briefly discuss the upper limits on the EDM $d_W$ of the
$W$-boson, as derived from the strict bounds on the electron and neutron EDM
\cite{ss,mq}. We follow the line of reasoning of Ref.~\cite{gkr} taking into
account, however, the arbitrary magnetic quadrupole moment of the $W$-boson omitted
in Refs~\cite{ss,mq,gkr}.

\begin{figure}[h]
\center
\includegraphics[scale=1.5]{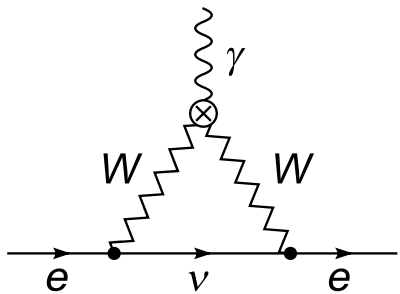}
\caption {The $d_W$ contribution to the electron EDM} \label{fig:4}
\end{figure}

We start the discussion with the $d_W$ contribution to the electron EDM. The effect
is described by diagram presented in Fig.~\ref{fig:4}.

In the previous papers the EDM interaction is described by the Lagrangian
\begin{equation} \label{d1}
L_W^{edm} = ie\widetilde{\kappa} W^\dag_{\mu}W_\nu\widetilde{F}^{\mu\nu}\,,
\end{equation}
so $d_W=e \widetilde{\kappa}/2m_W$.

Then the result for the $d_W$ contribution to the electron EDM is
\begin{equation} \label{del}
\Delta d_e = \,\frac{\alpha}{8\pi \sin^2
\theta_w}\,\frac{m_e}{m_W}\,\ln\frac{\Lambda^2}{m_W^2}\,d_W.
\end{equation}

Here $\Lambda$ is the cut-off parameter for the logarithmically divergent integral
over virtual momenta in the loop. Putting (perhaps, quite conservatively)
$\ln(\Lambda^2/m_W^2) \simeq 1$, one obtains with the experimental upper limit on
the electron EDM \cite{bcr} (see Table~\ref{table:1}), the following bound on the
dipole moment of $W$-boson:
\begin{equation} \label{We}
d_W/e \; \lesssim\; 2 \times 10^{-19} \; \rm{cm}
\end{equation}
(it corresponds to $\widetilde{\kappa} \lesssim 2 \times 10^{-3}$).

However, general consideration shows that a particle of spin one can have two $P$-
and $T$-odd electromagnetic moments \cite{kot}. In line with an electric dipole
moment, a vector particle can have a magnetic quadrupole moment (MQM) $Q$.

Therefore, the Lagrangian contains two independent structures, corresponding to
arbitrary EDM and MQM:
\begin{equation} \label{lagr1}
L_W=ie\widetilde{\kappa} W^\dagger_\mu W_\nu
\widetilde{F}^{\mu\nu}+\frac{ie\widetilde{\lambda}}{m^2_W}W^\dagger_{\mu\nu}
W^\nu{}_\rho \widetilde{F}^{\rho\mu}.
\end{equation}

Now $d_W$ and $Q_W$ are related to $\widetilde{\kappa}$ and $\widetilde{\lambda}$ as
follows:
\begin{equation}
d_W=\frac{e}{2m_W}\left(\widetilde{\kappa}+\widetilde{\lambda}\right),
\end{equation}

\begin{equation}
Q_W=-\frac{e}{m^2_W}\left(\widetilde{\kappa}-\widetilde{\lambda}\right),
\end{equation}
and the result for the $d_W$ contribution to the electron EDM is
\begin{equation}
\Delta d_e = \,\frac{\alpha}{8\pi \sin^2
\theta_w}\,\frac{m_e}{m_W}\,\frac{e}{2m_W}\left[\widetilde{\kappa}\ln\frac{\Lambda^2}{m_W^2}+\widetilde{\lambda}\right].
\end{equation}

Once again putting $\ln(\Lambda^2/m_W^2) \simeq 1$, one obtains the same bound
(\ref{del},\,\ref{We}) on the dipole moment of $W$-boson:
\begin{equation}
\Delta d_e \simeq \,\frac{\alpha}{8\pi \sin^2
\theta_w}\,\frac{m_e}{m_W}\,\frac{e}{2m_W}\,\left[\widetilde{\kappa}+\widetilde{\lambda}\right]=\frac{\alpha}{8\pi
\sin^2 \theta_w}\,\frac{m_e}{m_W}\,d_W,
\end{equation}
\begin{equation} \label{We1}
d_W/e \; \lesssim\; 2 \times 10^{-19} \; \rm{cm}\,.
\end{equation}

Of course, this result directly depends on the cut-off parameter $\Lambda$. If one
chooses $\Lambda \simeq 1$ TeV, then $\ln(\Lambda^2/m_W^2) \simeq 5$. However, in
this case one obtains the upper limit on some linear combination of
$\widetilde{\kappa}$ and $\widetilde{\lambda}$, but not on the $W$-boson EDM itself.

In the case of the $W$-boson contribution to the neutron EDM considered in
Ref.~\cite{gkr} (the EDM interaction therein is described by (\ref{d1})) the same
calculations, as those in the case of electron EDM, result in the following
expression for the discussed contribution to the neutron dipole moment:
\begin{equation} \label{dn}
\Delta d_n = \,\frac{\alpha}{8\pi \sin^2 \theta_w}\,\frac{m_n}{m_W}\left[g_0
\ln\frac{\Lambda^2}{m_W^2}\, + h_0\left(\ln\frac{\Lambda^2}{m_W^2} +
1\right)\right]d_W,
\end{equation}
where $g_0,\, h_0 \sim 1$.

For numerical estimate one can take
\[
g_0\ln\frac{\Lambda^2}{m_W^2}\, + h_0\left(\ln\frac{\Lambda^2}{m_W^2} + 1\right)
\sim 1\,,
\]  so that
\begin{equation}
\Delta d_n \sim \,\frac{\alpha}{8\pi \sin^2 \theta_w}\,\frac{m_n}{m_W}\,d_W\,\approx
\,\frac{\alpha}{2\pi}\,\frac{m_n}{m_W}\,d_W\,.
\end{equation}
Then, with the result of \cite{bak} for the neutron EDM (see Table~\ref{table:1}),
one arrives at the following quite strict upper limit on the $W$-boson dipole
moment:
\begin{equation}
d_W/e \; \lesssim\; 2 \times 10^{-21} \; \rm{cm}\,.
\end{equation}

Taking into account the second term in the Lagrangian (\ref{lagr1}) alters the
contribution of $d_W$ to the neutron EDM:
\begin{multline}
\Delta d_n = \,\frac{\alpha}{8\pi \sin^2
\theta_w}\,\frac{m_n}{m_W}\frac{e}{2m_W}\left[\widetilde{\kappa}\left\{g_0
\ln\frac{\Lambda^2}{m_W^2}+h_0\left(\ln\frac{\Lambda^2}{m_W^2} +
1\right)\right\}+\right.
\\ \left. +\widetilde{\lambda}\left\{g_0-h_0\left(\ln\frac{\Lambda^2}{m_W^2}-1\right)\right\}\right].
\end{multline}

For numerical estimate we put both expressions in curly brackets of the order of
unity, so
\begin{equation}
\Delta d_n \sim \,\frac{\alpha}{8\pi \sin^2
\theta_w}\,\frac{m_n}{m_W}\,\frac{e}{2m_W}\,\left[\widetilde{\kappa}+\widetilde{\lambda}\right]=
\,\frac{\alpha}{8\pi \sin^2 \theta_w}\,\frac{m_n}{m_W}\,d_W\,.
\end{equation}
Therefore the upper bound on the $W$-boson EDM does not change:
\begin{equation} \label{Wn}
d_W/e \; \lesssim\; 2 \times 10^{-21} \; \rm{cm}\,.
\end{equation}

The contribution of $d_W$ to the fermion EDM (see Fig.~\ref{fig:4}) is proportional
to the fermion mass. Therefore, due to the larger mass of the neutron, the limit on
$d_W$ (\ref{Wn}) following from $\Delta d_n$ is better than that (\ref{We1}) from
$\Delta d_e$ by two orders of magnitude, even though the experimental upper limit on
$d_e$ \cite{bcr} is better by an order of magnitude than that on $d_n$ \cite{bak}.

\subsection*{\normalsize 4. Conclusions}

We obtained the upper limits on the $CP$-violating TGC parameters by two different
methods: from the analysis of the total cross-sections of $e^+e^-\to W^+W^-$ and
from strict experimental bounds on the electron and neutron EDM. There is also third
way of doing this: by constructing $CP$-odd observables from the kinematic variables
in the reaction $e^+e^-\to W^+W^-$. Each of these methods has its advantages and
shortcomings. The analysis of the total cross-sections may give strict limits on
$\widetilde{\kappa}_V$, $\widetilde{\lambda}_V$ due to high statistics and hence
small statistical errors, but the bounds are obtained under the assumption that all
other anomalous TGC parameters vanish, thus they are not uniquely the limits on the
$CP$-violating couplings. On the contrary, $CP$-odd observables are sensitive to
these couplings, but statistics is less than in the previous case. The limits
derived from the bounds on the electron and neutron EDM are also uniquely the limits
on $\widetilde{\kappa}_V$, $\widetilde{\lambda}_V$ and they are much better than the
limits obtained from the $e^+e^-$ annihilation. However, they are of rather
qualitative nature because they depend on the uncertain cut-off parameter $\Lambda$.

\subsection*{\normalsize Acknowledgements}

We are grateful to I.B.\,Khriplovich for many helpful discussions and suggestions as
well as for critical reading of the article.

The work was supported in part by the Russian Foundation for Basic Research through
Grant No.~08-02-00960-a, by the Federal Program "Personnel of Innovational Russia"
through Grant No.~14.740.11.0082, and by Dmitry Zimin's Dynasty Foundation.

\renewcommand{\refname}


{\normalsize References}
\begin{thebibliography}\\

\bibitem{gp}
G.J.\,Gounaris, C.G.\,Papadopoulos, Eur.\,Phys.\,J.~\textbf{C 2}, 365 (1998).

\bibitem{lvy}
A.A.\,Likhoded, G.\,Valencia and O.P.\,Yushchenko, Phys.\,Rev.~\textbf{D 57}, 2974
(1998).

\bibitem{ckk}
D.\,Choudhury, J.\,Kalinowski and A.\,Kulesza, Phys.\,Lett.~\textbf{B 457}, 193
(1999).

\bibitem{dnn}
M.\,Diehl, O.\,Nachtmann and F.\,Nagel, Eur.\,Phys.\,J.~\textbf{C 27}, 375 (2003);
Eur.\,Phys.\,J.~\textbf{C 32}, 17 (2003).

\bibitem{opal}
The OPAL Collaboration, Eur.\,Phys.\,J.~\textbf{C 19}, 229 (2001).

\bibitem{del}
DELPHI Collaboration, Eur.\,Phys.\,J.~\textbf{C 54}, 345 (2008).

\bibitem{bcr}
B.C.\,Regan, E.D.\,Commins, C.J.\,Schmidt, D.\,DeMille,
Phys.\,Rev.\,Lett.~\textbf{88}, 071805 (2002).

\bibitem{bak}
C.A.\,Baker {\em et al}., Phys.\,Rev.\,Lett.~\textbf{97}, 131801 (2006).

\bibitem{ds}
V.F.\,Dmitriev, R.A.\,Sen'kov, Phys.\,Rev.\,Lett.~\textbf{91}, 212303 (2003).

\bibitem{for}
W.C.\,Griffith, M.D.\,Swallows, T.H.\,Loftus, M.V.\,Romalis, B.R.\,Heckel,
E.N.\,Fortson, Phys.\,Rev.\,Lett.~\textbf{102}, 101601 (2009).

\bibitem{ben}
G.W.\,Bennett {\em et al}., Phys.\,Rev.~\textbf{D 80}, 052008 (2009).

\bibitem{hag}
K.\,Hagiwara, R.D.\,Peccei, D.\,Zeppenfeld, K.\,Hikasa, Nucl.\,Phys.~\textbf{B 282},
253 (1987).

\bibitem{gou}
G.\,Gounaris {\em et al}., CERN 96-01, 525, arXiv:hep-ph/9601233.

\bibitem{quei}
A.\,Queijeiro, Phys.\,Rev.~\textbf{D 39}, 3507 (1989).

\bibitem{br}
A.E.\,Blinov, A.S.\,Rudenko,  Nucl.\,Phys.~\textbf{B} (Proc. Suppl.) \textbf{189},
257 (2009).

\bibitem{bil}
M.S.\,Bilenky, J.L.\,Kneur, F.M.\,Renard, D.\,Schildknecht, Nucl.\,Phys.~\textbf{B
409}, 22 (1993).

\bibitem{hwz}
K.\,Hagiwara, J.\,Woodside, D.\,Zeppenfeld, Phys.\,Rev.~\textbf{D 41}, 2113 (1990).

\bibitem{LEPEWWG}
LEPEWWG group, hep-ex/0612034.

\bibitem{ss} F.\,Salzman, G.\,Salzman, Phys.\,Lett.~\textbf{15}, 91 (1965);
Nuovo Cimento~\textbf{A 41}, 443 (1966).

\bibitem{mq} F.J.\,Marciano, A.\,Queijeiro, Phys.\,Rev.~\textbf{D 33}, 3449 (1986).

\bibitem{gkr} A.G.\,Grozin, I.B.\,Khriplovich, A.S.\,Rudenko, Nucl.\,Phys.~\textbf{B 821}, 285 (2009).

\bibitem{kot} I.Yu.\,Kobzarev, L.B.\,Okun', M.V.\,Terentyev, Sov.\,Phys.\,JETP\,Lett.~\textbf{2}, 289 (1965).

\end{thebibliography}
\end{document}